\documentclass[journal]{IEEEtran}
\ifCLASSINFOpdf
\else
\fi
\usepackage{amsmath}
\usepackage{amsfonts}
\usepackage{subcaption}
\usepackage{url}
\usepackage{multirow}
\usepackage{enumitem}
\usepackage{algpseudocode}
\usepackage{algorithm}

\usepackage{xcolor}

\usepackage{multicol}
\usepackage{graphicx}
\usepackage{cite}
\usepackage{balance}

\begin{document}
%
\title{ManiGen: A Manifold Aided Black-box Generator of Adversarial Examples}
%
%
%

\author{Guanxiong~Liu, Issa~Khalil, Abdallah Khreishah, Abdulelah Algosaibi, Adel Aldalbahi, Mohammed Alaneem, Abdulaziz Alhumam, and Mohammed Anan
\thanks{G. Liu and A. Khreishah are with the Department
of Electrical and Computer Engineering, New Jersey Institute of Technology, Newark, NJ, 07102 USA e-mail: (gl236@njit.edu).}
\thanks{A. Algosaibi, A. Aldalbahi, M. Alaneem, and A. Alhumam are with King Faisal University.}
\thanks{M. Anan is with Alfaisal University.}}

\maketitle

\begin{abstract}

Machine learning models, especially neural network (NN) classifiers, have acceptable performance and accuracy that leads to their wide adoption in different aspects of our daily lives. The underlying assumption is that these models are generated and used in attack free scenarios. However, it has been shown that neural network based classifiers are vulnerable to adversarial examples. Adversarial examples are inputs with special perturbations that are ignored by human eyes while can mislead NN classifiers. Most of the existing methods for generating such perturbations require a certain level of knowledge about the target classifier, which makes them not very practical. For example, some generators require knowledge of pre-softmax logits while others utilize prediction scores. 

In this paper, we design a practical \textbf{black-box} adversarial example generator, dubbed \textbf{ManiGen}. ManiGen does not require any knowledge of the inner state of the target classifier. It generates adversarial examples by searching along the manifold, which is a concise representation of input data. Through extensive set of experiments on different datasets, we show that (1) adversarial examples generated by ManiGen can mislead standalone classifiers by being as successful as the state-of-the-art white-box generator, \textit{Carlini}, and (2) adversarial examples generated by ManiGen can more effectively attack classifiers with state-of-the-art defenses.

\end{abstract}

\begin{IEEEkeywords}
Adversarial Examples, Machine Learning, Neural Network, Manifold
\end{IEEEkeywords}

\section{Introduction}

\begin{figure}[tb]
\centering
\begin{minipage}[c]{.45\textwidth}
\centering
    \includegraphics[width=\linewidth]{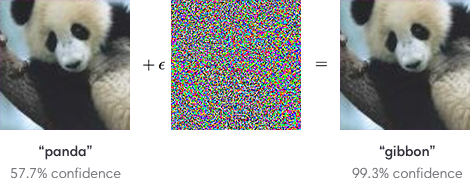}
    \caption{Fast Gradient Sign Example \cite{goodfellow2014explaining}}
    \label{fig:fgm-example}
\end{minipage}
\vspace{-5mm}
\end{figure}

Due to the surprisingly good representation power of complex distributions, neural network models are widely used in many applications including natural language processing, computer vision and cyber security. For example, in cyber security, neural network (NN) classifiers are used for spam filtering, phishing detection as well as face recognition \cite{rowley1998neural} \cite{abu2007comparison}. However, the training and usage of NN classifiers are based on an underlying assumption that the environment is attack free. Therefore, such classifiers fail when adversarial examples are presented to them. Adversarial examples were first introduced in 2013 by Szegedy et. al \cite{szegedy2013intriguing} in the context of image classification. They show that adversarial examples can be generated by adding specially designed perturbations to original images. As shown in Figure~\ref{fig:fgm-example}, such perturbations are visually insignificant to human eyes, but strong enough to mislead the classifier to output an incorrect result. Yet, more scary, the results show that the success rate of such attack against standalone classifiers can reach 100\%. That is, all the generated adversarial examples are missclassified.

The introduction of adversarial examples presents a serious threat to the applicability and adoption of neural network models, especially in critical and sensitive applications such as banking and security \cite{papernot2018sok}. For example, adversarial examples can mislead a neural network model that identifies digits written on a check, which may result in incorrect cashed amounts. On another example, a specially designed image may be used to bypass a security check that utilizes neural networks. This serious threat inspires a new line of research to understand the mechanisms exploited to generate adversarial examples, in order to develop appropriate defensive mechanisms.

One of the early attempts to understand adversarial examples was performed by Szegedy et al. \cite{szegedy2013intriguing}, in which they estimated that the root cause behind the successful generation of adversarial examples against NN classifiers is the highly non-linear property of neural networks. However, a follow on work by Goodfellow et. al \cite{goodfellow2014explaining} shows that adversarial examples can be generated by the fast gradient sign method. This simply means that even very linear neural network models are also vulnerable to adversarial examples. Therefore, more researchers believe that adversarial examples against NN classifiers are possible due to the linearity of the models, not its non-linearity, and due to its inability to handle high dimensional data.

Current generators of adversarial examples can be categorized into white-box and black-box versions. The white-box generators usually formulate an optimization problem which utilizes information from target classifier's prediction or inner states. By solving the optimization problem, the adversarial examples are generated \cite{goodfellow2014explaining, kurakin2016adversarial}. In contrast to white-box generators, black-box generators assume the target classifier is intransparent. Therefore, black-box generators always try to utilize indirect information to generate adversarial examples \cite{papernot2016practical, chen2017zoo}. From the practical perspective, the black-box generators are more useful in the real-world scenario. However, the current black-box generators are far from perfect. The trasferability-based black-box generator \cite{papernot2016practical} has lower success rate due to the difference between target and substitute classifiers. And, the zeroth order black-box generator \cite{chen2017zoo} requires the prediction confidence on all classes from the target classifier.

In this paper, we first propose a practical way to generate adversarial examples by searching along the manifold of the training data. The manifold is a hyperplane formed by low dimensional data points from the original high dimensional space. To search for adversarial examples, our approach, dubbed \textbf{ManiGen}, solves an optimization problem which can search along the manifold using the Gradient Descent Algorithm \cite{zeiler2012adadelta} \cite{kingma2014adam}. To the best of our knowledge, ManiGen is the first \textbf{black-box} approach that utilizes the manifold of the training data to generate adversarial examples. That is, \textit{ManiGen can generate adversarial examples with zero knowledge of the structure and inner state from the target classifier.}

Compared with the existing black-box generators, our proposed ManiGen has the following advantages. Firstly, it does not rely on the transferability of adversarial examples between target and substitute classifiers \cite{papernot2016practical}. Secondly, it does not requires extra information other than prediction results from the target classifier \cite{chen2017zoo}.

We conduct experiments to compare the adversarial examples generated by ManiGen and those generated by \textit{Carlini}, which is a state-of-the-art white-box generator. The results show that ManiGen adversarial examples can mislead standalone classifiers with 100\% success rate which is at the same level as Carlini. More importantly, the results also show that the adversarial examples generated by ManiGen are more threatening to classifiers with defenses.

Our contributions in this paper can be summarized as follows: 
\begin{itemize}[leftmargin=*]
    \item We design a \textbf{black-box} approach for generating adversarial examples against NN classifier. More specifically, our approach, dubbed \textbf{ManiGen}, is distinguished by being model agnostic, that is, it generates adversarial examples without demanding any information about the inner states of the target classifier. Instead, ManiGen utilizes the autoencoder based approach to search adversarial examples along the manifold of the training data.
    
    \item Compared with the existing black-box generators \cite{papernot2016practical, chen2017zoo}, ManiGen outperforms them in two aspects. (1) It does not depend on the transferability of adversarial examples between target and substitute classifiers. (2) It does not require information other than prediction results from target classifier.
    
    \item We show through extensive set of experiments that ManiGen adversarial examples have 100\% success rate in misleading standalone NN classifiers. More importantly, adversarial examples generated from ManiGen are more threatening to classifiers with defenses.
\end{itemize}

The rest of the paper is organized as follows. Section \ref{sec:background} presents background and related work. The designs of ManiGen are presented in Sections \ref{sec:attack}. Section \ref{sec:setting}, presents the test-bed design and the experimental settings. The evaluation results are presented in Section \ref{sec:results} and Section \ref{sec:conclusion} concludes the paper.

\section{Background and Related Work} \label{sec:background}

In this section, we introduce high level background knowledge about NN classifiers, autoencoder and manifolds for  better understanding of the concepts presented in this work. We also, provide relevant references for further information about each topic.

\subsection{Neural Network Classifier}\label{subsec:nnc}

Neural network (NN) were introduced by several biological and computer science researchers in early 1900s but did not have the due attention because of its high computation cost. However, it lately gain more traction due to the considerable increase in computing power, which makes them practical in many machine learning applications including computer vision, pattern recognition, cyber security, and natural language processing \cite{deng2014deep}. In this paper, we consider the popular use case of NN in classification applications. After being trained with labeled set of data, a NN classifier can learn to classify new input instances into one of a number of predefined set of classes.

\textcolor{black}{
The general structure of a NN is shown in Figure \ref{fig:nn-example}. It could be seen as a collection of layers and each layer contains a certain number of neurons. The connections between neurons have different weights, $\omega$, and the value of a neuron is a non-linear transformation of the weighted sum of neurons connected to it. This non-linear transformation is called activation function and defined by user. In NN classifier, the weighted sum value in the final layer neurons are also called pre-softmax logits, $Z$, since the activation function of final layer is usually softmax function which is defined as $f(x_{i}) = \frac{e^{x_{i}}}{\sum e^{x_{i}}}$.
}

From the system level point of view, a NN classifier is equivalent to a highly non-linear function which maps input data to a prediction. Based on the difference between the predictions and the corresponding ground truth, a loss function is formulated as a function of $\Theta$, which is denoted as $loss_{\Theta}$. During the training process, the gradient decent algorithm \cite{zeiler2012adadelta} \cite{kingma2014adam} is used to iteratively minimize the value of the loss function. Finally, the famous backward propagation algorithm is used \cite{rumelhart1985learning} to update the gradients throughout the whole NN.

\subsection{Autoencoder and Manifold}

Autoencoder is an unsupervised NN model. It is designed to learn a representation, which is called code, of a set of data through unsupervised learning. Typically, the autoencoder is used to learn a low dimensional encoding of a high dimensional data. For example, a normal $32 \times 32$ RGB image can be represented as a 3D matrix of $32\times32\times3=3072$ dimensional data. The autoencoder takes the 3D matrix representation of the image as input and generates a 128 dimensional vector representation of the image.

\begin{figure*}[tb]
\centering
\begin{minipage}[c]{.3\textwidth}
\centering
    \includegraphics[width=\linewidth]{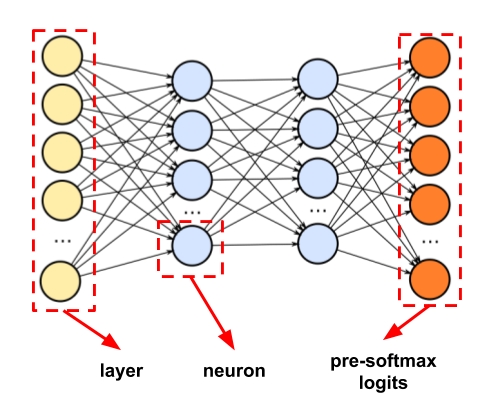}
    \caption{Neural Network Example}
    \label{fig:nn-example}
\end{minipage}
\begin{minipage}[c]{.3\textwidth}
\centering
    \includegraphics[width=\linewidth]{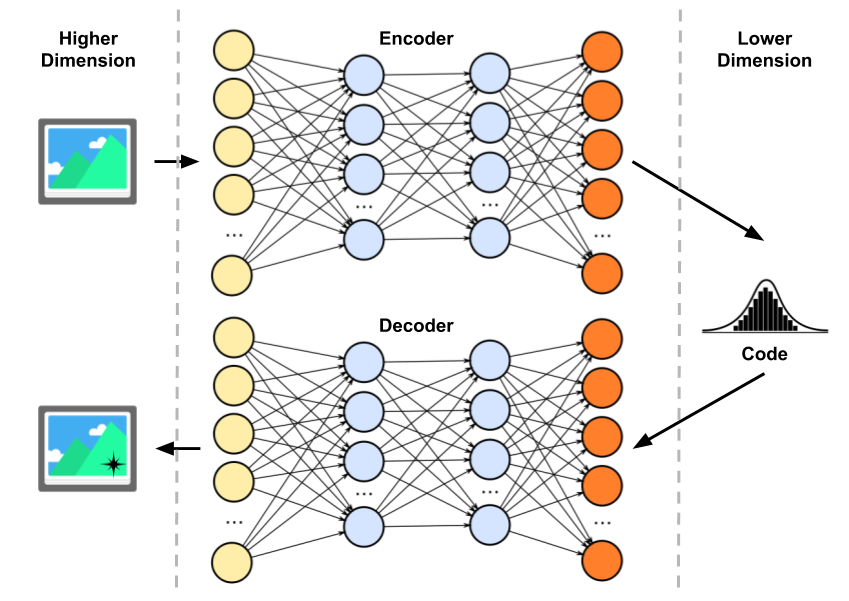}
    \caption{Autoencoder \cite{wiki:Autoencoder}}
    \label{fig:ae-structure}
\end{minipage}
\begin{minipage}[c]{.3\textwidth}
\centering
    \includegraphics[width=\linewidth]{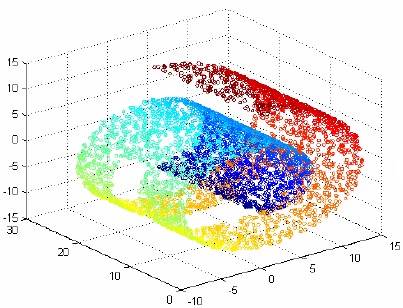}
    \caption{Manifold Example}
    \label{fig:manifold-example}
\end{minipage}
\end{figure*}

The general structure of the autoencoder is composed of an encoder and a decoder as depicted in Figure~\ref{fig:ae-structure}. Through the connected layers inside the encoder, input data is squeezed and mapped to a code, which is usually a lower dimensional data. Then, the decoder takes the code and expands it back to the original size. During the training, the difference between input data and the output from the decoder is calculated and used as a loss function value. By compressing and extracting input data with a lower loss function value, the autoencoder learns how to efficiently map high dimensional input data to a lower dimensional code. 

In machine learning community, researchers believe that the relevant data for a certain task lies on a hyperplane, called \textbf{Manifold}, in the original high dimensional space \cite{narayanan2010sample}. By flattening the manifold, we can get a lower dimensional representation of the original data points. The autoencoder is one of the most popular tools to learn the manifold and generate lower dimensional codes from input data. The distance between any two codes can be used to reflect the similarity of their corresponding data points. Since the code is in a much lower dimensional space, this distance is more representative and more reliable than the distance calculated from the original high dimensional space. \textcolor{black}{Based on the property of autoencoder, it is widely used in computer vision and natural language processing for semantic analysis \cite{goodfellow2016deep}.}

\subsection{Generators of Adversarial Examples}

The generators of adversarial examples against NN classifiers are generally categorized into white-box and black-box based approaches. White-box based generators are usually easier to build, but not very practical as they require access to the target NN classifier \cite{kurakin2016adversarial} \cite{carlini2016towards}. On the other hand, black-box based generators are model agnostic as they can operate without any knowledge of the inner working of the target NN classifier \cite{meng2017magnet}. Black-box based generators are more practical, however, they are much harder to design and implement. Therefore, the majority of existing adversarial example generators are white-box based ones. \textcolor{black}{In the following, we highlight the design approach of the state-of-the-art adversarial example generators.}

\begin{figure*}
\centering
\begin{minipage}[c]{.3\textwidth}
\centering
    \includegraphics[width=\linewidth]{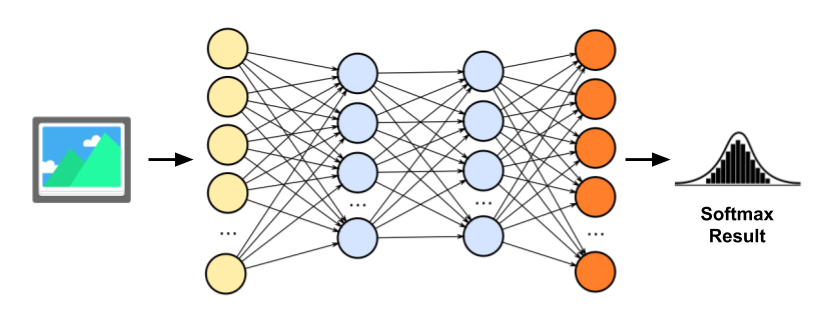}
    \caption{White-box Approximation}
    \label{fig:white-box}
\end{minipage}
\begin{minipage}[c]{.3\textwidth}
\centering
    \includegraphics[width=\linewidth]{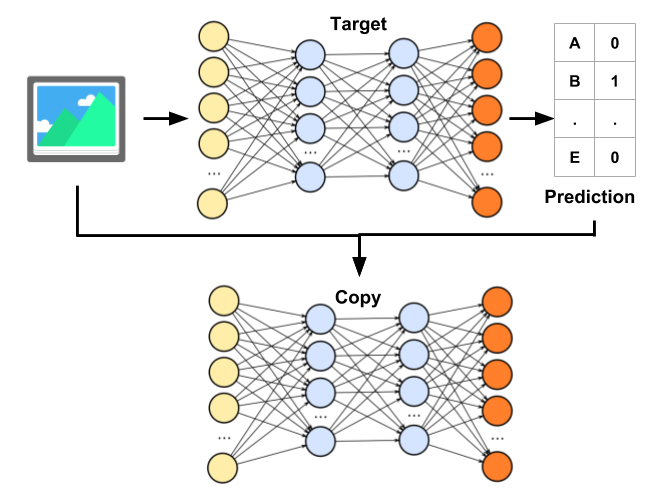}
    \caption{Black-box Approximation}
    \label{fig:black-box-others}
\end{minipage}
\begin{minipage}[c]{.3\textwidth}
\centering
    \includegraphics[width=\linewidth]{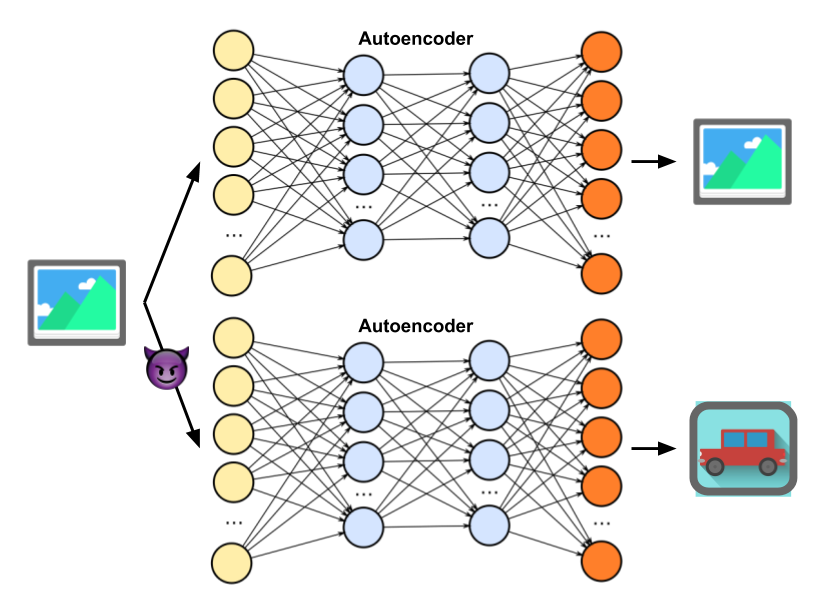}
    \caption{ManiGen Approximation}
    \label{fig:black-box-our}
\end{minipage}
\end{figure*}

\textcolor{black}{
In the general format, the adversarial example generator could be formulated as an optimization problem which searches a small neighbor area of the original image for the existence of adversarial examples. Assume an original image could be denoted as $x$ and the perturbation we made to it could be represented by $\delta$, the process of searching for adversarial examples could be formulated as follows:
\begin{equation*}
\begin{aligned}
    & \underset{\delta}{\text{minimize}}
    & & \mathcal{D} (x, \delta) \\
    & \text{subject to}
    & & \mathcal{C} (x, \delta) \neq t \\
    & 
    & & \mathcal{F} (x + \delta) \in [0,1]^{m}
\end{aligned}
\end{equation*}
By solving this optimization problem, a special perturbation $\delta$, which could minimize the visual difference from original image that is denoted as $\mathcal{D}$, will be found. There are several different choices for the $\mathcal{D}$ function which include $L_{2}$, $L_{0}$ and $L_{\infty}$. In this paper, to make the $\mathcal{D}$ function fully differentiable, ManiGen choose $L_{2}$ distance which is a normal choice in many other works.
}

\textcolor{black}{
In the optimization formulation, we called the first constraint as \textbf{``effective constraint"} while the second constraint as \textbf{``validation constraint"}. The current perturbation is strong enough to fool the classifier $\mathcal{C}$ if and only if $\mathcal{C} (x + \delta) \neq t$ where $t$ is the true label of original image. The $\mathcal{F}$ function ensures that the generated adversarial example is still a valid image.
}

\textcolor{black}{
From the original formulation of this optimization problem, the "effective constraint" $\mathcal{C}$ could be naturally explained as the classification function. However, due to the truth that modern classifier is highly non-linear, it is hard to directly solve the optimization problem in the original format and different methods have different approximation of this "effective constraint". For white-box attack, the softmax result or the loss function based on it are usually used since these attack could get access to inner information from target classifier as Figure \ref{fig:white-box}.
}

\textbf{Gradient Sign Method} is introduced by Goodfellow et. al in \cite{goodfellow2014explaining} as a white-box adversarial example image generator against NN image classifiers. It simply generates adversarial examples from original images by adding a small value ($\epsilon$) from each pixel in the same direction of the gradients of the loss function. From the previous formulation, this method utilizes $\epsilon$ as step size to make sure the adversarial examples are visually similar. At the same time, it tries to maximize the loss function, which represents the similarity between prediction and ground truth, in order to fool the classifier. Therefore, it tries to solve the following optimization problem.
\begin{equation*}
\begin{aligned}
    & \underset{\delta}{\text{maximize}}
    & & \text{loss}_{f} (x, \delta, t) \\
    & \text{subject to}
    & & \mathcal{F} (x + \delta) \in [0,1]^{m}
\end{aligned}
\end{equation*}

The early implementation of this formulation is the Fast Gradient Sign Method (FGSM) \cite{goodfellow2014explaining} which only applies gradients once. A more recent work in \cite{kurakin2016adversarial} introduces the Basic Iterative Method (BIM) which repeats the gradient sign method in multiple smaller steps to generate more threatening adversarial examples.

\textbf{Carlini Attack Generator} is a set of white-box attacks designed by Nicholas Carlini \cite{carlini2016towards}. The generators are carefully designed to utilize pre-softmax logits to calculate gradients. This makes it effective in misleading classifiers which try to hide their final prediction scores from adversaries. 
\textcolor{black}{
The approach in this work could be formulated as follows with $e = \underset{i \neq t}{\text{max}} (Z(x, \delta)_{i}) - Z(x, \delta)_{t})$. Here, the $Z$ function is defined in Section \ref{subsec:nnc}.
\begin{equation*}
\begin{aligned}
    & \underset{\delta}{\text{minimize}}
    & & \mathcal{D} (x, \delta) + c \times \text{max} (0,~e) \\
    & \text{subject to}
    & & \mathcal{F} (x + \delta) \in [0,1]^{m}
\end{aligned}
\end{equation*}
}
This method is recognized as a standard for white-box generators since it can efficiently utilize pre-softmax results from target classifiers to generate adversarial examples, which are shown to defeat many of the state-of-the-art defensive methods. However, similar to the previous method, it is not very practical due to the required access to the pre-sofmax results.

\textbf{Black-box Generators} are more practical than the white-box generators. Existing works show two ways of generating black-box adversarial examples. As shown in Figure \ref{fig:black-box-others}, one way of doing so is to train a substitute which approximates the decision boundary of the target classifier \cite{papernot2016practical}. Once the substitute is ready, any white-box attack method could be used to generate adversarial examples. However, the difference between substitute and target classifiers degenerates the success rate of generating adversarial examples. Beside this, other researchers propose a zeroth order method which empirically approximates the gradients \cite{chen2017zoo}. It changes the value of each pixel in a small range and measures the corresponding changes in prediction confidence. The problem is that this approach cannot be applied when only prediction results are provided.

\subsection{Adversarial Example Defensive Methods}

Based on literature review, we select two different representative defense methods of adversarial examples. The first approach aims at performing dimensional reduction and transformation of input data, while the other approach focuses on utilizing some adversarial examples to retrain the target classifiers. In the following, We introduce one representative example from each of the these approaches:

\textbf{Dimensional Reduction and Transformation} is introduced by Meng et. al \cite{meng2017magnet}. In this work, the authors design two types of functional components which are called detector and reformer. Adversarial examples are either rejected by the detector or reformed to clean up the adversarial perturbations. The detector can be designed based on an autoencoder structure or based on the probability divergence concept \cite{meng2017magnet}. On the other hand, the reformer is simply an autoencoder based detector. The reformer is designed to recognize and remove small adversarial perturbation in input samples. On the other hand, the detector is designed to recognize adversarial examples with strong perturbations that bypass the reformer. The main drawback of this method is that the reformer can only work efficiently with simple images and fails to efficiently handle perturbations in complex images.

\textbf{Adversarial Training} is based on a simple idea that treats adversarial examples as blind spots of the original training data \cite{xu2016automatically}. Through retaining by samples of adversarial examples, the classifier learns new features from adversarial examples and generalizes its prediction to account for such perturbations. By far, adversarial training is one of the most efficient ways to mitigate adversarial examples. However, adversarial training based defenses \cite{goodfellow2014explaining,liu2019using,liu2019gandef,liu2019zk,liu2020using} require access to adversarial examples during the training process, which makes it effective only against known adversarial examples and may fail with new ones.

\section{Manifold based Attack Model} \label{sec:attack}


In this section, we introduce our approach for generating adversarial examples. 
\textcolor{black}{Based on the general format introduced in Section \ref{sec:background}, we show our approximation of the optimization problem. After that, we introduce the constraints of the optimization problem.}

\subsection{Objective Function}

\textcolor{black}{Compared with methods introduced previously, our approach aims at generating black-box adversarial examples. Therefore, our method has no access to the inner information from target classifier. Therefore, the searching processing in black-box fashion is non-trivial since we can not change the searching direction based on reaction of the classifier through gradient decent method.}

Based on this situation, we introduce the autoencoder and manifold learning from unsupervised learning area into our approach. From many existing works in the manifold learning area \cite{narayanan2010sample} \cite{lawrence2012unifying}, people noticed that the relevant data of an AI task are always near or on a manifold which has a much lower dimension than the full sample space and Figure~\ref{fig:manifold-example} could be seen as an example. Moreover, by projecting data samples onto this manifold, we are trying to keep both linear and non-linear data structure information.

Intuitively speaking, in the image classification application, the original image space could be seen as "visual space" while its projection space on the manifold could be seen as "semantic space". To combine the manifold learning and let it guide the searching of adversarial example, we made the following assumption:

\textbf{\textit{Assumption :} We assume that there exists a set of examples which are close to the original sample in the ``visual space" but relatively far away from it in the ``semantic space". When the distance is close enough in ``visual space" and far enough in "semantic space", these examples are the adversarial examples to the original sample.}


Based on our previous assumption, we have two methods to approach the "effective constraint" $\mathcal{C}$ as:
\begin{equation*}
\begin{aligned}
    & & \mathcal{C} \neq t \rightarrow ||\mathcal{AE} (x, \delta) - \mathcal{AE} (x)||_{2} \geq d_{\text{visual}} \\
    & \text{OR} \\
    & & \mathcal{C} \neq t \rightarrow ||\mathcal{E} (x, \delta) - \mathcal{E} (x)||_{2} \geq d_{\text{semantic}}
\end{aligned}
\end{equation*}
where $\mathcal{AE}$ represents the autoencoder's non-linear transformation and $\mathcal{E}$ represents the encoder's non-linear transformation. In this paper, all of our adversarial examples are generated by utilizing the first implementation since the generated images are more similar to original sample. The possible reason could be related to that L2 distance metric in ``visual space" and ``semantic space" are not suitable to be added together directly. However, we get this understanding from experiments directly and a detailed study of these two implementations could be in future works.

\textcolor{black}{
Now, our autoencoder based approximation of the optimization problem without constraint could be summarized as follows and Figure \ref{fig:black-box-our} gives an intuitive visualization of utilizing autoencoder for our approximation.
\begin{equation*}
\begin{aligned}
    & \underset{\delta}{\text{minimize}}
    & & \mathcal{D} (x, \delta) - c \times ||\mathcal{AE} (x, \delta) - \mathcal{AE} (x)||_{2} \\
    & \text{subject to}
    & & \mathcal{F} (x + \delta) \in [0,1]^{m}
\end{aligned}
\end{equation*}
}

\subsection{Constraint}

After the transformation of ``effective constraint", the only constraint left in the original optimization problem is the ``validation constraint". By applying this constraint, we want to make sure the generated adversarial example is a valid image. However, in order to make the searching more efficient, we want to find a proper way to handle this problem. To achieve this goal, we first denote that $\delta \in \mathbb{R}^{m}$. Then, we could naturally define the transform function as:
\begin{equation*}
\begin{aligned}
    & \mathcal{F} (x, \delta) = \frac{1}{2} \tanh(\arctan(2 \times (x - \frac{1}{2})) + \delta) + \frac{1}{2}
\end{aligned}
\end{equation*}

In our implementation, we use $\arctan(2 \times (x - \frac{1}{2}) \times 0.99999)$ instead to avoid the situation of calculating $\arctan(1)$ or $\arctan(-1)$. With the help of this transformation, we could get rid of directly checking the ``validation constraint" which may decrease the searching efficiency. Moreover, with this transformation, generating adversarial examples could be achieved by solving an unconstrained optimization problem with no limitation on the algorithm.

\section{Experimental Settings} \label{sec:setting}

This section presents the experimental ecosystem we use to evaluate the confusion capabilities of adversarial examples generated by our proposed ManiGen. The main components of our test-bed include: (1) Pre-processing module, (2) Attack module, (3) Classification module, and (4) Defense module. The attack and defense modules serve as containers in which different attack and defense models are plugged and unplugged based on specific experimental settings. The input data used by the test-bed is scaled and then separated into training and testing sets. The pre-processed data and, in some cases, the original data flow into the attack module, which generates adversarial examples and interacts with the classification module to refine them. The refined adversarial examples are then passed to the defensive module to evaluate its effectiveness in correctly classifying the examples. 

We explain different attack models in earlier sections and in the following sections, we elaborate on the remaining components of the test-bed, after presenting the datasets utilized.

\subsection{Datasets}

We use the following datasets to conduct our experiments:
\begin{itemize}
    \item MNIST: Contains a total of 70K, $28 \times 28$, gray scale labeled images of handwritten digits. 
    \item CIFAR10: Contains a total of 60K, $32 \times 32$, RGB labeled images of animals and vehicles. 
    \item STL10: Contains a total of 6K, $96 \times 96$, RGB labeled images of animals and vehicles.
\end{itemize}
The images in each dataset are evenly labeled into 10 different classes.

\subsection{Pre-processing}

Pre-processing involves the following operations:
\begin{itemize}[leftmargin=*]
    \item Scaling: Gray scale images use one integer to represent each of its pixels, while RGB images use three different integers (each between 0 and 255) to represent each of its pixels. To simplify the process of finding adversarial examples and to be consistent with the related work, scaling is used to map pixel representations from discrete integers in the range $\mathbb{Z}_{[0,255]}$ into real numbers in the range $\mathbb{R}_{[0,1]}$.
    \item Separation: This operation is used to split each input dataset into two groups: training-dataset and testing-dataset. The training dataset is used to train the supervised machine learning modules including, autoencoders, classifiers, and defensive modules, while the testing dataset is used by the attack modules to guide the generation of adversarial examples, as well as for evaluating the defensive modules. The 70K MNIST images are randomly separated into 60K training and 10K testing images, the 60K CIFAR10 images are randomly separated into 50K training and 10K testing images, and the 6K STL10 images are randomly separated into 5K training and 1K testing images.
    \item Autoencoder: As mentioned earlier, the autoencoder is utilized in our black-box based adversarial example generator. It is trained with training datasets and used to guide the search along the manifold to generate our adversarial examples. We utilize two different deep convolutional autoencoder structures, which are shown in Table~\ref{table:autoencoder-structure}. For MNIST, we encode each original image from $784~(28 \times 28)$ dimension into $128~(4 \times 4 \times 8)$ dimension in \textit{semantic space}, while in CIFAR10 and STL10, we encode each image from $3072~(32 \times 32 \times 3)$ and $27648~(96 \times 96 \times 3)$ dimensions into $256~(4 \times 4 \times 16)$ dimension. The training settings of these autoencoders are presented in Table~\ref{table:autoencoder-training}. 
\end{itemize}

\begin{table}[h]
    \begin{center}
    \begin{tabular}{c | c | c | c} 
    \hline
    \multicolumn{4}{c}{Autoencoder} \\[0.5ex] 
    \hline \hline
    Layer & Parameter & Padding & Activation \\
    \hline 
    \multicolumn{4}{c}{Encoder} \\
    \hline
    Convolution & $3 \times 3 \times X_{1}$ & Same & ReLU \\ 
    \hline
    Pooling & $2 \times 2$ & Same & - \\
    \hline
    Convolution & $3 \times 3 \times X_{2}$ & Same & ReLU \\ 
    \hline
    Pooling & $2 \times 2$ & Same & - \\
    \hline
    Convolution & $3 \times 3 \times X_{2}$ & Same & ReLU \\ 
    \hline
    Pooling & $2 \times 2$ & Same & - \\
    \hline 
    \multicolumn{4}{c}{Decoder} \\
    \hline
    Convolution & $3 \times 3 \times X_{2}$ & Same & ReLU \\ 
    \hline
    Up Sampling & $2 \times 2$ & - & - \\
    \hline
    Convolution & $3 \times 3 \times X_{2}$ & Same & ReLU \\ 
    \hline
    Up Sampling & $2 \times 2$ & - & - \\
    \hline
    Convolution & $3 \times 3 \times X_{1}$ & Same & ReLU \\ 
    \hline
    Up Sampling & $2 \times 2$ & - & - \\
    \hline
    Convolution & $3 \times 3 \times X_{3}$ & Same & Sigmoid \\ 
    \end{tabular}
    \end{center}
    \caption{Autoencoder Structure (For MNIST, $X_{1}=16$, $X_{2}=8$, $X_{3}=1$ and using Max Pooling. For CIFAR10 and STL10, $X_{1}=32$, $X_{2}=16$, $X_{3}=3$ and using Average Pooling.)}
    \label{table:autoencoder-structure}
\vspace{-3mm}
\end{table}
\begin{table}[h]
    \begin{center}
    \begin{tabular}{c | c | c} 
    \hline
    \multicolumn{3}{c}{Autoencoder Training} \\[0.5ex] 
    \hline \hline
    Setting & MNIST & CIFAR10 \& STL10 \\
    \hline
    Optimizer & Adam & Adam \\
    Learning Rate & 0.01 & 0.01 \\
    Loss Function & Binary Cross-Entropy & Mean Square Error \\
    Batch Size & 128 & 256 \\
    Shuffle & Yes & Yes \\
    Epoch & 50 & 100 \\
    \end{tabular}
    \end{center}
    \caption{Autoencoder Training Setting}
    \label{table:autoencoder-training}
\vspace{-3mm}
\end{table}

\subsection{Classification Module}

This module implements standalone classifier and is mainly used by the attack component of the test-bed to guide the process of adversarial example generation. After an adversarial example is generated, the attack generator forwards it to the target classifier to check if it succeeds in misleading that classifier. The result of the test is fed back to the adversarial example generator, which uses it to refine the next adversarial examples. We note that the output of the target classifier is customized based on the specific adversarial example generator. For example, ManiGen example generator expects a binary output that indicates whether the prediction is right or wrong. On the other hand, the Carlini generator expects the output to include the final logits $Z$ from the classifier, which contain the inner information from the last NN layer.

In our test-bed evaluation, we select a different classifier for each dataset with performance that matches the relevant benchmarks \cite{benchmark-list}. The structure for the classifier used with the MNIST dataset is shown in  Tabel~\ref{table:mnist-classifier-structure}, while the structure for the allCNN classifier \cite{springenberg2014striving} used with the CIFAR10 dataset is shown in Table~\ref{table:cifar-classifier2-structure}. For the STL10 dataset classifier, we apply transfer learning based on VGG16 model \cite{simonyan2014very} trained on ImageNet dataset \cite{imagenet_cvpr09}. The training settings of each of the three classifiers are shown in Table~\ref{table:classifier-training}. It is worth to note that softmax logits (or pre-softmax logits $Z$) is required by existing white-box and black-box adversarial example generators. As a more practical black-box generator, our ManiGen does not require this information.
\begin{table}[h]
    \begin{center}
    \begin{tabular}{c | c | c | c} 
    \hline
    \multicolumn{4}{c}{MNIST - Classifier} \\[0.5ex] 
    \hline \hline
    Layer & Parameter & Padding & Activation \\
    \hline
    Convolution & $3 \times 3 \times 16$ & Same & ReLU \\ 
    \hline
    Max. Pooling & $2 \times 2$ & Same & - \\
    \hline
    Convolution & $3 \times 3 \times 8$ & Same & ReLU \\ 
    \hline
    Max. Pooling & $2 \times 2$ & Same & - \\
    \hline
    Convolution & $3 \times 3 \times 8$ & Same & ReLU \\ 
    \hline
    Max. Pooling & $2 \times 2$ & Same & - \\
    \hline
    Dense       & $128$        & - & ReLU \\
    \hline
    Dense       & $10$         & - & Softmax \\
    \end{tabular}
    \end{center}
    \caption{MNIST Classifier Structure}
    \label{table:mnist-classifier-structure}
\vspace{-3mm}
\end{table}
\begin{table}[h]
    \begin{center}
    \begin{tabular}{c | c | c | c} 
    \hline
    \multicolumn{4}{c}{CIFAR10 - Classifier} \\[0.5ex] 
    \hline \hline
    Layer & Parameter & Padding & Activation \\
    \hline
    Convolution & $3 \times 3 \times 96$ & Same & ReLU \\ 
    \hline
    Convolution & $3 \times 3 \times 96$ & Same & ReLU \\ 
    \hline
    Convolution & $3 \times 3 \times 96$ & Same & ReLU \\ 
    \hline
    Max Pooling & $2 \times 2$ & Same & - \\
    \hline
    Dropout     & 0.5 & - & - \\
    \hline
    Convolution & $3 \times 3 \times 192$ & Same & ReLU \\ 
    \hline
    Convolution & $3 \times 3 \times 192$ & Same & ReLU \\ 
    \hline
    Convolution & $3 \times 3 \times 192$ & Same & ReLU \\ 
    \hline
    Max Pooling & $2 \times 2$ & Same & - \\
    \hline
    Dropout     & 0.5 & - & - \\
    \hline
    Convolution & $3 \times 3 \times 192$ & Same & ReLU \\ 
    \hline
    Convolution & $1 \times 1 \times 192$ & Same & ReLU \\ 
    \hline
    Convolution & $1 \times 1 \times 10$ & Same & ReLU \\ 
    \hline
    Global Avg. Pooling & - & - & - \\
    \hline
    Dense       & 10 & - & Softmax \\
    \end{tabular}
    \end{center}
    \caption{CIFAR10 allCNN Classifier Structure}
    \label{table:cifar-classifier2-structure}
\vspace{-3mm}
\end{table}
\begin{table}[h]
    \begin{center}
    \begin{tabular}{c | c | c | c} 
    \hline
    \multicolumn{4}{c}{Classifier - Training} \\[0.5ex] 
    \hline \hline
    Setting & MNIST & CIFAR10 & STL10 \\
    \hline
    Optimizer & Adam & SGD & Adadelta\\
    Learning Rate & 0.01 & 0.01 & 1.0 \\
    Loss Function & Cross Entropy & Cross Entropy & Cross Entropy \\
    Batch Size & 128 & 32 & 32 \\
    Shuffle & Yes & Yes & Yes \\
    Width Shift & - & 0.2 & - \\
    Height Shift & - & 0.2 & - \\
    Horizontal Flip & - & True & - \\
    Epoch & 100 & 350 & 50 \\
    \end{tabular}
    \end{center}
    \caption{Classifier Training Setting}
    \label{table:classifier-training}
\vspace{-3mm}
\end{table}

\subsection{Attack Module}

The attack module implements two adversarial example generators, the Carlini and the ManiGen. Each of the generators interacts with the appropriate standalone classifier, according to the dataset under consideration, to refine its generated adversarial examples. The pre-processed data and, in some cases, the original data flow into the attack module, which iteratively solves an optimization problem to generate adversarial examples. Recall that the Carlini method is a white-box generator which requires the pre-softmax logits that contains the inner information of the target classifier. While, for the black-box ManiGen, it only need to know that the classifier is making right or wrong prediction. For both adversarial example generators (Carlini and ManiGen), we only run the algorithm for the same predefined number of iterations. The final generated examples are used as adversarial examples for evaluation.

\subsection{Defensive Module}

In the defensive module, we implement two approaches: (i) MagNet \cite{meng2017magnet}, and (ii) traditional adversarial training method, which we call AdvDef \cite{kurakin2016adversarial}. In the implemntaion of MagNet, we follow its original designe as presented in \cite{meng2017magnet}. For AdvDef, we utilize the same structures as those of the standalone classifiers mentioned earlier. 

Adversarial examples generated by the attack module are passed to the defensive module to evaluate its capability to mislead the classifier with defense. During the evaluation, the only measurement considered in this work is the test accuracy. However, for different kinds of input examples, our definition of test accuracy is different. If the input examples are original examples, the test accuracy follows its traditional definition which can be formulated as:
\begin{align}
    \text{Test Accuracy} \equiv \frac{\text{\# of correctly classified input examples}}{\text{total \# of input examples}}
\end{align}
When the input examples are adversarial examples, both making correct prediction and rejecting the input are considered as a success from the defender's perspective. Now, the definition of test accuracy is as follows:
\begin{align}
\begin{split}
    &\text{Test Accuracy} \equiv \\
    &\frac{\text{\# of correctly classified and rejected input examples}}{\text{total \# of input examples}}
\end{split}
\end{align}

\section{Experimental Results} \label{sec:results}

\begin{figure*}[tb]
\centering
\begin{minipage}[c]{.3\textwidth}
\centering
    \includegraphics[width=\linewidth]{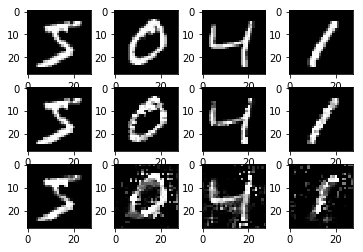}
    \label{fig:mnist-vis-1}
\end{minipage}
\begin{minipage}[c]{.3\textwidth}
\centering
    \includegraphics[width=\linewidth]{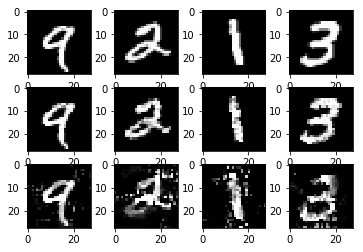}
    \label{fig:mnist-vis-2}
\end{minipage}
\begin{minipage}[c]{.3\textwidth}
\centering
    \includegraphics[width=\linewidth]{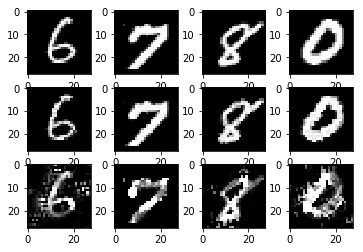}
    \label{fig:mnist-vis-3}
\end{minipage}
\caption{Visualization of Original and Adversarial Examples in MNIST}
\label{fig:mnist-vis}
\end{figure*}
\begin{figure*}[h!]
\centering
\begin{minipage}[c]{.3\textwidth}
\centering
    \includegraphics[width=\linewidth]{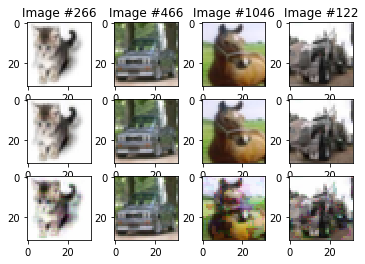}
    \label{fig:cifar-vis-1}
\end{minipage}
\begin{minipage}[c]{.3\textwidth}
\centering
    \includegraphics[width=\linewidth]{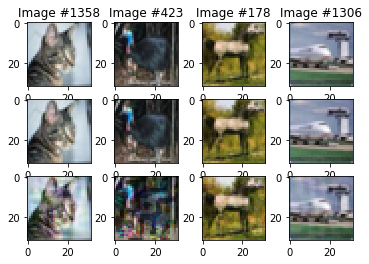}
    \label{fig:cifar-vis-2}
\end{minipage}
\begin{minipage}[c]{.3\textwidth}
\centering
    \includegraphics[width=\linewidth]{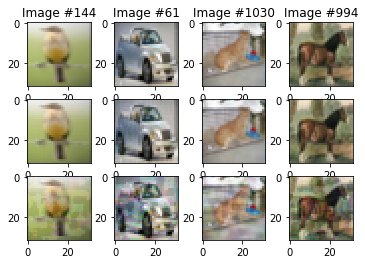}
    \label{fig:cifar-vis-3}
\end{minipage}
\caption{Visualization of Original and Adversarial Examples in CIFAR10}
\label{fig:cifar-vis}
\end{figure*}
\begin{figure*}[h!]
\centering
\begin{minipage}[c]{.3\textwidth}
\centering
    \includegraphics[width=\linewidth]{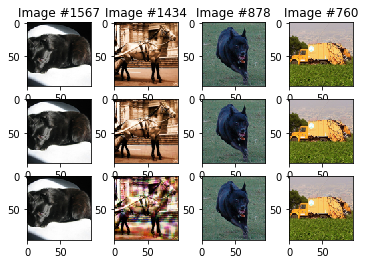}
    \label{fig:stl-vis-1}
\end{minipage}
\begin{minipage}[c]{.3\textwidth}
\centering
    \includegraphics[width=\linewidth]{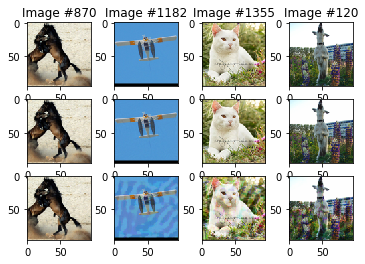}
    \label{fig:stl-vis-2}
\end{minipage}
\begin{minipage}[c]{.3\textwidth}
\centering
    \includegraphics[width=\linewidth]{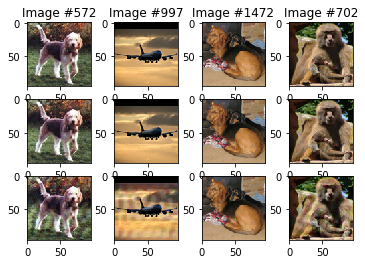}
    \label{fig:stl-vis-3}
\end{minipage}
\caption{Visualization of Original and Adversarial Examples in STL10}
\label{fig:stl-vis}
\end{figure*}

In this section, we present comparative evaluation results of our adversarial example generators, ManiGen. We first visualize and compare adversarial examples generated by ManiGen from MNIST, CIFAR10, and STL10 datasets, with those generated by Carlini and the corresponding original images. Then, we evaluate the capability of ManiGen adversarial examples in misleading classifiers with or without defense.

During the evaluation, we select the Carlini generator as the reference. It is worth to notice that the Carlini method is a white-box generator and it upper bounds the existing black-box generators. For the defenses, we consider both unsupervised (MagNet) as well as supervised (AdvDef) approaches. For the practical purpose in black-box setting, we do not consider adaptive attack against MagNet.


\subsection{Visualizations of Adversarial Examples}

In this subsection, we visually compare ManiGen and Carlini adversarial examples to their corresponding original images. Figures~\ref{fig:mnist-vis}-\ref{fig:stl-vis} show random samples from the three datasets, MNIST, CIFAR10, and STL10, respectively. In each figure, the first row presents the original images, the second row presents the corresponding Carlini examples, and the last row presents the corresponding ManiGen examples. 

Figure~\ref{fig:mnist-vis} presents randomly selected set of images from MNIST, which cover all the ten different decimal digits. The figure shows that ManiGen digits 5, 9, 7, and 8 are visually close to the corresponding Carlini ones. On the other hand, the ManiGen digits 4, 1, 3, and 6 contain more highlighted (i.e., white) noise pixels. Additionally, ManiGen digits 2 and 0 have both extra dark and white noise pixels compared to the Carlini and the original images. Therefore, some Carlini examples (e.g., digit 0) looks slightly closer to the original images than the corresponding ManiGen examples. However, presented alone, these figures are still easy to be recognized by humans. There are two aspects of reasons which include (1) the MNIST images are simple and (2) the adversarial perturbation generated by ManiGen does not change the semantic meaning of the image.

When we move to the CIFAR10 dataset in Figure \ref{fig:cifar-vis}, the difference between the Carlini and ManiGen adversarial examples is becoming smaller. In CIFAR10, the Carlini and ManiGen adversarial examples are almost identical in many images (\#226, \#466, \#122, \#1358, \#178, \#1306, and \#994). Among others, the ManiGen adversarial examples contain stronger perturbation. However, such adversarial perturbation is meaningless and cannot mislead human's prediction.

Lastly, the visualization in STL10 dataset (Figure \ref{fig:stl-vis}) reflects the aforementioned trend again. Only in few of these images (\#1434, \#1182, and \#1355), the ManiGen adversarial examples have stronger perturbation. Similar to before, these perturbations are more like a meaningless noise which cannot mislead human's prediction. While in all other images, the Carlini and ManiGen adversarial examples are quite similar to each other.

Generally speaking, the slightly reduced visual quality of some of ManiGen examples compared to Carlini is due to the black-box design of ManiGen. That is, ManiGen does not utilize any information from the classifier to guide the search process for adversarial examples. In other words, ManiGen trades slight decrease in visual quality of some adversarial examples with the more practical black-box design approach. Recall that ManiGen substitutes the lack of information about the target classifier by searching across the approximated manifold. Therefore, ManiGen needs to add, in some cases, relatively larger perturbations to generate successful attack examples. However, through our extensive experiments on the three datasets (MNIST, CIFAR10, and STL10), we observe that such perturbations are semantically meaningless to human vision system especially when images get more complex. Overall, both the ManiGen and Carlini adversarial examples are hard to fool human vision system while extremely easy to mislead classifier's prediction.

\begin{table}[tb]
    \centering
    \begin{tabular}{ c c | c c c }
         \hline \hline
         \multirow{2}{*}{Dataset} & \multirow{2}{*}{Example} & \multicolumn{3}{c}{Classifier} \\
         & & Standalone & MagNet & AdvDef \\
         \hline
         \multirow{3}{*}{MNIST} & Original & 99.7\% & 98.1\% & 98.4\% \\
         & Carlini & 0\% & 100\% & 99.5\% \\
         & ManiGen & 0\% & 97.5\% & 64.5\% \\
         \hline
         \multirow{3}{*}{CIFAR10} & Original & 89.5\% & 81.5\% & 86.2\% \\
         & Carlini & 0\% & 54.7\% & 88.5\% \\
         & ManiGen & 0\% & 46.9\% & 73\% \\
         \hline
         \multirow{3}{*}{STL10} & Original & 87.7\% & 48.9\% & 81.7\% \\
         & Carlini & 0\% & 51.5\% & 62\% \\
         & ManiGen & 1.5\% & 37\% & 58.5\% \\
         \hline \hline
    \end{tabular}
    \caption{Summary of Test Accuracy}
    \label{table:summary-res}
\end{table}

\subsection{Confusion Capabilities of Adversarial Examples}

In this subsection, we evaluate the ability of ManiGen and Carlini adversarial examples to mislead classifiers. Recall that, an example misleads a classifier if it causes the classifier to output a wrong classification class. We build an experimental labeled dataset which consists of 384 randomly selected original images from each of the three datasets (a total of 1152 images), the corresponding Carlini adversarial examples, and the corresponding ManiGen adversarial examples. Then, we classify the images using (1) standalone classifier, (2) classifier with MagNet, and (3) adversarially trained classifier, AdvDef. The detailed structures and the training settings of these classifiers are explained earlier in Section \ref{sec:setting}. The results of measured test accuracy are summarized in Table \ref{table:summary-res}

In the first column of Table \ref{table:summary-res}, we start with the standalone classifier's test accuracy on original examples. In all three datasets, the standalone classifier can achieve over 87\% of test accuracy which means that these classifiers are well trained under the attack free assumption. As we expected, the test accuracy on ManiGen or Carlini adversarial examples has a significant degeneration down to almost 0\%. This clearly shows that ManiGen and Carlini examples from all the datasets have 100\% success rate in misleading the standalone classifier.

In the next column, we repeat these measurements on the same input examples. The difference is that the classifier is protected by the MagNet. We can see that the classifier performs much better than the standalone one on all the datasets. More importantly, the results show that ManiGen examples are more resistant to the defense (MagNet) than Carlini ones. As shown in Table \ref{table:summary-res}, the MagNet's test accuracy on ManiGen examples are always lower than that on Carlini examples. More importantly, the difference increases with the increase in the complexity of the dataset from MNIST to STL10. It is worth to mention that the MagNet itself is hard to be scaled to complex dataset. Its test accuracy on original examples decreases significantly compared with standalone classifier in the STL10 dataset.

In the last column of Table \ref{table:summary-res}, we present the test accuracy of the AdvDef, on different input examples. Compared with MagNet, AdvDef has better scalability and is wildly accepted as defense method against adversarial examples. However, AdvDef is far from perfect especially in complex dataset. For example in STL10 dataset, the test accuracy on Carlini examples is only 62\% compared with over 81\% test accuracy on original examples. If we compare the ManiGen and Carlini examples, we can see that using ManiGen can mislead AdvDef as successful as using Carlini. Actually, adversarial examples generated by ManiGen are more threatening (lower test accuracy).

To summarize, we see from the evaluation results that (1) both ManiGen and Carlini can mislead standalone classifier with nearly 100\% success rate and (2) when the classifier is equipped with defense (MagNet or AdvDef), ManiGen can generate more threatening adversarial examples than Carlini.

\section{Conclusion} \label{sec:conclusion}

In this paper, we introduce a new adversarial example generator, dubbed \textbf{ManiGen}, that searches adversarial examples along the manifold in a black-box fashion. Compared with existing black-box generators, our approach has higher success rate in misleading standalone classifier. Moreover, it does not require classifier's prediction confidence. 

Through extensive experiments, we evaluate the confusion capabilities of ManiGen and compare it with the white-box generator, Carlini. The results show that it misleads standalone machine learning models as successful as the Carlini on different datasets. More seriously, it can mislead the classifier with different defenses (MagNet and AdvDef) more effectively than the Carlini.

\balance

\ifCLASSOPTIONcaptionsoff
  \newpage
\fi



\bibliographystyle{IEEEtran}
\bibliography{reference}

\begin{thebibliography}{10}
\providecommand{\url}[1]{#1}
\csname url@samestyle\endcsname
\providecommand{\newblock}{\relax}
\providecommand{\bibinfo}[2]{#2}
\providecommand{\BIBentrySTDinterwordspacing}{\spaceskip=0pt\relax}
\providecommand{\BIBentryALTinterwordstretchfactor}{4}
\providecommand{\BIBentryALTinterwordspacing}{\spaceskip=\fontdimen2\font plus
\BIBentryALTinterwordstretchfactor\fontdimen3\font minus
  \fontdimen4\font\relax}
\providecommand{\BIBforeignlanguage}[2]{{%
\expandafter\ifx\csname l@#1\endcsname\relax
\typeout{** WARNING: IEEEtran.bst: No hyphenation pattern has been}%
\typeout{** loaded for the language `#1'. Using the pattern for}%
\typeout{** the default language instead.}%
\else
\language=\csname l@#1\endcsname
\fi
#2}}
\providecommand{\BIBdecl}{\relax}
\BIBdecl

\bibitem{rowley1998neural}
H.~A. Rowley, S.~Baluja, and T.~Kanade, ``Neural network-based face
  detection,'' \emph{IEEE Transactions on pattern analysis and machine
  intelligence}, vol.~20, no.~1, pp. 23--38, 1998.

\bibitem{abu2007comparison}
S.~Abu-Nimeh, D.~Nappa, X.~Wang, and S.~Nair, ``A comparison of machine
  learning techniques for phishing detection,'' in \emph{Proceedings of the
  anti-phishing working groups 2nd annual eCrime researchers summit}.\hskip 1em
  plus 0.5em minus 0.4em\relax ACM, 2007, pp. 60--69.

\bibitem{szegedy2013intriguing}
C.~Szegedy, W.~Zaremba, I.~Sutskever, J.~Bruna, D.~Erhan, I.~Goodfellow, and
  R.~Fergus, ``Intriguing properties of neural networks,'' \emph{International
  Conference on Learning Representations}, 2014.

\bibitem{goodfellow2014explaining}
I.~J. Goodfellow, J.~Shlens, and C.~Szegedy, ``Explaining and harnessing
  adversarial examples,'' \emph{International Conference on Learning
  Representations}, 2015.

\bibitem{papernot2018sok}
N.~Papernot, P.~McDaniel, A.~Sinha, and M.~P.~Wellman, ``Sok: Security and
  privacy in machine learning,'' \emph{IEEE European Symposium on Security and
  Privacy}, 2018.

\bibitem{kurakin2016adversarial}
A.~Kurakin, I.~Goodfellow, and S.~Bengio, ``Adversarial machine learning at
  scale,'' \emph{International Conference on Learning Representations}, 2017.

\bibitem{papernot2016practical}
N.~Papernot, P.~McDaniel, I.~Goodfellow, S.~Jha, Z.~B. Celik, and A.~Swami,
  ``Practical black-box attacks against deep learning systems using adversarial
  examples,'' \emph{ACM Asia Conference on Computer and Communications
  Security}, 2017.

\bibitem{chen2017zoo}
P.-Y. Chen, H.~Zhang, Y.~Sharma, J.~Yi, and C.-J. Hsieh, ``Zoo: Zeroth order
  optimization based black-box attacks to deep neural networks without training
  substitute models,'' in \emph{Proceedings of the 10th ACM Workshop on
  Artificial Intelligence and Security}, 2017, pp. 15--26.

\bibitem{zeiler2012adadelta}
M.~D. Zeiler, ``Adadelta: an adaptive learning rate method,'' \emph{arXiv
  preprint arXiv:1212.5701}, 2012.

\bibitem{kingma2014adam}
D.~P. Kingma and J.~Ba, ``Adam: A method for stochastic optimization,''
  \emph{International Conference on Learning Representations}, 2015.

\bibitem{deng2014deep}
L.~Deng, D.~Yu \emph{et~al.}, ``Deep learning: methods and applications,''
  \emph{Foundations and Trends{\textregistered} in Signal Processing}, vol.~7,
  no. 3--4, pp. 197--387, 2014.

\bibitem{rumelhart1985learning}
D.~E. Rumelhart, G.~E. Hinton, and R.~J. Williams, ``Learning internal
  representations by error propagation,'' California Univ San Diego La Jolla
  Inst for Cognitive Science, Tech. Rep., 1985.

\bibitem{wiki:Autoencoder}
\BIBentryALTinterwordspacing
W.~contributors, ``Autoencoder,'' 2018, [Online; accessed 06-April-2018].
  [Online]. Available: \url{https://en.wikipedia.org/wiki/Autoencoder}
\BIBentrySTDinterwordspacing

\bibitem{narayanan2010sample}
H.~Narayanan and S.~Mitter, ``Sample complexity of testing the manifold
  hypothesis,'' in \emph{Advances in Neural Information Processing Systems},
  2010, pp. 1786--1794.

\bibitem{goodfellow2016deep}
I.~Goodfellow, Y.~Bengio, A.~Courville, and Y.~Bengio, \emph{Deep
  learning}.\hskip 1em plus 0.5em minus 0.4em\relax MIT press Cambridge, 2016,
  vol.~1.

\bibitem{carlini2016towards}
N.~Carlini and D.~Wagner, ``Towards evaluating the robustness of neural
  networks,'' pp. 39--57, 2017.

\bibitem{meng2017magnet}
D.~Meng and H.~Chen, ``Magnet: a two-pronged defense against adversarial
  examples,'' pp. 135--147, 2017.

\bibitem{xu2016automatically}
W.~Xu, Y.~Qi, and D.~Evans, ``Automatically evading classifiers,'' in
  \emph{Proceedings of the 2016 Network and Distributed Systems Symposium},
  2016.

\bibitem{liu2019using}
G.~Liu, I.~Khalil, and A.~Khreishah, ``Using intuition from empirical
  properties to simplify adversarial training defense,'' in \emph{2019 49th
  Annual IEEE/IFIP International Conference on Dependable Systems and Networks
  Workshops (DSN-W)}.\hskip 1em plus 0.5em minus 0.4em\relax IEEE, 2019, pp.
  58--61.

\bibitem{liu2019gandef}
------, ``Gandef: A gan based adversarial training defense for neural network
  classifier,'' in \emph{IFIP International Conference on ICT Systems Security
  and Privacy Protection}.\hskip 1em plus 0.5em minus 0.4em\relax Springer,
  2019, pp. 19--32.

\bibitem{liu2019zk}
------, ``Zk-gandef: A gan based zero knowledge adversarial training defense
  for neural networks,'' in \emph{2019 49th Annual IEEE/IFIP International
  Conference on Dependable Systems and Networks (DSN)}.\hskip 1em plus 0.5em
  minus 0.4em\relax IEEE, 2019, pp. 64--75.

\bibitem{liu2020using}
------, ``Using single-step adversarial training to defend iterative
  adversarial examples,'' \emph{arXiv preprint arXiv:2002.09632}, 2020.

\bibitem{lawrence2012unifying}
N.~D. Lawrence, ``A unifying probabilistic perspective for spectral
  dimensionality reduction: Insights and new models,'' \emph{Journal of Machine
  Learning Research}, vol.~13, no. May, pp. 1609--1638, 2012.

\bibitem{benchmark-list}
\BIBentryALTinterwordspacing
R.~Benenson, ``Classification datasets results,'' 2018, [Online; accessed
  06-April-2018]. [Online]. Available:
  \url{http://rodrigob.github.io/are_we_there_yet/build/classification_datasets_results.html}
\BIBentrySTDinterwordspacing

\bibitem{springenberg2014striving}
J.~T. Springenberg, A.~Dosovitskiy, T.~Brox, and M.~Riedmiller, ``Striving for
  simplicity: The all convolutional net,'' \emph{International Conference on
  Learning Representations}, 2017.

\bibitem{simonyan2014very}
K.~Simonyan and A.~Zisserman, ``Very deep convolutional networks for
  large-scale image recognition,'' \emph{International Conference on Learning
  Representations}, 2015.

\bibitem{imagenet_cvpr09}
J.~Deng, W.~Dong, R.~Socher, L.-J. Li, K.~Li, and L.~Fei-Fei, ``{ImageNet: A
  Large-Scale Hierarchical Image Database},'' in \emph{CVPR09}, 2009.

\end{thebibliography}
\end{document}